\documentclass[aps,prd,onecolumn,showpacs,amsmath,floatfix]{revtex4}
\usepackage{subfigure}
\usepackage[dvips]{graphicx}
\usepackage{mathptmx}      % use Times fonts if available on your TeX system
\usepackage{bm}

\begin{document}

\def\be{\begin{equation}}
\def\ee{\end{equation}}
\def\bea{\begin{eqnarray}}
\def\eea{\end{eqnarray}}
\def\rra{\right\rangle}
\def\lla{\left\langle}
\def\eps{\epsilon}
\def\sgm{\Sigma^-}
\def\la{\Lambda}
\def\pv{\bm{p}}
\def\kv{\bm{k}}
\def\zv{\bm{0}}
\def\bc{B=90\;\rm MeV\!/fm^3}
\def\ms{M_\odot}

\title{Hybrid protoneutron stars with the MIT bag model}

\author{O. E. Nicotra, M. Baldo, G. F. Burgio, and H.-J. Schulze}
\affiliation{
Dipartimento di Fisica e Astronomia, Universit\`a di Catania and \\
INFN, Sezione di Catania, Via Santa Sofia 64, 95123 Catania, Italy}

\date{\today}

\begin{abstract}
We study the hadron-quark phase transition in the interior of protoneutron
stars.
For the hadronic sector, we use a microscopic equation of state
involving nucleons and hyperons 
derived within the finite-temperature Brueckner-Bethe-Goldstone many-body
theory, with realistic two-body and three-body forces.
For the description of quark matter, we employ the MIT bag model
both with a constant and a density-dependent bag parameter.
We calculate the structure of protostars with the equation of state
comprising both phases and find maximum masses below 1.6 solar masses.
Metastable heavy hybrid protostars are not found.
\end{abstract}

\maketitle

%==============================================================================
\section{Introduction}

It is generally believed that a neutron star (NS) is formed as a result
of the gravitational collapse of a massive star ($M \gtrsim 8\ms$)
in a type-II supernova \cite{shapiro,bethe}. 
Just after the core bounce, a protoneutron star (PNS) 
is formed, a very hot and lepton-rich object, 
where neutrinos are temporarily trapped. 
The following evolution of the PNS is dominated by neutrino diffusion,
which results first in deleptonization and subsequently in cooling.
The star stabilizes at practically zero temperature, and no trapped 
neutrinos are left.

The dynamical transformation of a PNS into a NS 
could be strongly influenced by a phase transition in the central region 
of the star.
Calculations of PNS structure, based on a microscopic nucleonic
equation of state (EOS), indicate that for the heaviest PNS, close
to the maximum mass (about two solar masses), the central particle density
reaches values larger than $1/\rm fm^{3}$.
In this density range the nucleon cores (dimension $\approx 0.5\;\rm fm$)
start to touch each other, and it is likely that quark 
degrees of freedom will play a role.

In a previous article \cite{proto} we have studied static properties 
of PNS assuming that nucleons, hyperons, and leptons are present 
in stellar matter. Our calculations are based on the EOS
derived within the Brueckner-Bethe-Goldstone (BBG) theory of nuclear matter, 
extended to finite temperature.
We have found that for purely nucleonic stars,
both thermal effects and neutrino trapping slightly soften the EOS, thus  
reducing the value of the maximum mass. 
If hyperons are included, neutrino
trapping shifts their onset to larger baryon density, 
and instead stiffens the EOS. 
This could lead to metastability during the subsequent
evolution of the PNS to the late neutrino-free stage.

In this work we extend the previous calculations, and take into account 
a possible hadron-quark phase transition. 
In fact, as in the case of cold NS, 
the addition of hyperons demands for the inclusion of quark 
degrees of freedom in order to obtain a maximum mass larger than the 
observational lower limit. 
For this purpose we use the BBG EOS 
for describing the hadronic phase and the MIT bag model at finite
temperature for the quark matter (QM) phase. 
We employ both a constant and a density-dependent bag parameter $B$. 
We find that the presence of QM 
increases the value of the maximum mass of a PNS, and 
stabilizes it at about 1.5--1.6 $\ms$, no matter the value of the 
temperature. 
In contrast to purely hyperonic stars, neutrino trapping in hybrid stars 
does not increase the maximum mass and does not allow metastable states.

The paper is organized as follows. In section \ref{s:bhf} we review
the determination of the baryonic EOS comprising nucleons and
hyperons in the finite-temperature Brueckner-Hartree-Fock approach.
Section \ref{s:qm} concerns the QM EOS according to the
MIT bag model. In section \ref{s:res} we present the results
regarding PNS structure combining the baryonic and
QM EOS for beta-stable nuclear matter. Section \ref{s:end}
contains our conclusions.

%==============================================================================
\section{Brueckner theory}
\label{s:bhf}

%------------------------------------------------------------------------------
\subsection{EOS of nuclear matter at finite temperature}

In the recent years, the BBG perturbative theory 
for nuclear matter has made much
progress, since its convergence has been firmly established \cite{thl}
and it has been extended in a fully microscopic and self-consistent way 
to the hyperonic sector \cite{hypmat,hypns,ns97}. 
Only few microscopic calculations of the nuclear EOS at finite temperature
are so far available. 
The first semi-microscopic investigation of the finite-temperature EOS
was performed in Ref.~\cite{fried}.
The results predict a Van der Waals behavior for symmetric nuclear matter, 
which leads to a liquid-gas 
phase transition with a critical temperature \hbox{$T_C \approx$ 18--20 MeV.} 
Later, Brueckner calculations \cite{lej,nic}
and chiral perturbation theory at finite temperature \cite{chi}
confirmed these findings with very similar values of $T_C$.
The Van der Waals behavior was also found in the finite-temperature
relativistic Dirac-Brueckner calculations of \cite{malfliet,huber}, 
although at a lower temperature.

The formalism which is closest to the BBG expansion, and actually
reduces to it in the zero-temperature limit, is the one formulated
in \cite{bloch}. 
In this approach the
essential ingredient is the two-body scattering matrix $K$, which,
along with the single-particle potential $U$, satisfies the
self-consistent equations
\bea
  \langle k_1 k_2 | K(W) | k_3 k_4 \rangle
 &=& \langle k_1 k_2 | V | k_3 k_4 \rangle
\nonumber\\&& \hskip-32mm %27 
 +\; \mathrm{Re}\!\sum_{k_3' k_4'}
 \langle k_1 k_2 | V | k_3' k_4' \rangle
 \frac{ [1\!-n(k_3')] [1\!-n(k_4')]}{W - E_{k_3'} - E_{k_4'} + i\epsilon }
%\nonumber\\&& \hskip-12mm \times
 \langle k_3' k_4' | K(W) | k_3 k_4 \rangle
\nonumber\\&&
\label{eq:kkk}
\eea
and
\be
 U(k_1) = \sum_{k_2} n(k_2) \langle k_1 k_2 | K(W) | k_1 k_2 \rangle_A \:,
\label{eq:ueq}
\ee
where $k_i$ generally denote momentum, spin, and isospin. 
Here $V$ is the two-body interaction, and we choose the Argonne $V_{18}$
nucleon-nucleon potential \cite{v18}.
$W = E_{k_1} + E_{k_2}$ represents the starting energy, 
$E_k = k^2\!/2m + U(k)$ the single-particle energy,
and $n(k)$ is a Fermi distribution. 
At $T = 0$ Eq.~(\ref{eq:kkk}) coincides with the Brueckner equation for
the $K$ matrix at zero temperature.

For given nucleon densities and temperature, 
Eqs.~(\ref{eq:kkk}) and (\ref{eq:ueq}) have to be
solved self-consistently along with the equations for the
%grand-canonical potential density $\omega$ 
densities, $\rho_i=\sum_k n_i(k)$,
and the free energy density, 
which has the following simplified expression
\be
 f = \sum_{i=n,p} \left[ \sum_{k} n_i(k)
 \left( \frac{k^2}{2m_i} + \frac{1}{2}U_i(k) \right) - Ts_i \right] \:,
 \label{e:fb}
\ee
where
\be
 s_i = - \sum_{k} \Big( n_i(k) \ln n_i(k) + [1-n_i(k)] \ln [1-n_i(k)] \Big)
\ee
is the entropy density for component $i$ treated as a free gas with
spectrum $E_i(k)$. 

In deriving Eq.~(\ref{e:fb}), we have introduced the so-called
{\em Frozen Correlations Approximation}, 
i.e., the correlations at $T\neq 0$ are assumed to be essentially 
the same as at $T=0$. 
This means that the single-particle potential $U_i(k)$ for the component $i$ 
can be approximated by the one calculated at $T=0$. 
This allows to save computational time and simplify the numerical procedure. 
It turns out that the assumed independence is valid to a good accuracy, 
at least for not too high temperature
(see Ref.~\cite{nic}, Fig.~12).
For a more extensive discussion of this topic,
the reader is referred to Ref.~\cite{book}, and references therein.

In our many-body approach, we have also introduced three-body forces
(TBF) among nucleons, in order to reproduce correctly the nuclear
matter saturation point 
$\rho_0 \approx 0.17~\mathrm{fm}^{-3}$, $E/A \approx -16$ MeV
and obtain values of incompressibility and symmetry energy at saturation
point compatible with those extracted from phenomenology \cite{myers}.
Since a complete microscopic theory of TBF is not
available yet, we have adopted the phenomenological Urbana model \cite{uix},
which in the BBG approach is reduced 
to a density-dependent two-body force by averaging over the position 
of the third particle, assuming that the probability of having two 
particles at a given distance is modified according to the 
two-body correlation function \cite{nic,bbb}. 

The fast rise of the nucleon chemical potentials with density in
NS cores \cite{glenhyp,gle}
may trigger the appearance of strange baryonic species, i.e., hyperons.
For this purpose we have extended the BBG approach in order to include the
$\Sigma^-$ and $\Lambda$ hyperons 
in a fully self-consistent way \cite{hypmat,hypns}.
In our work we have used the Nijmegen soft-core nucleon-hyperon (NH) 
potentials NSC89 \cite{maessen}, 
and neglected the hyperon-hyperon (HH) interactions, 
since so far no reliable HH potentials are available.
Recently also calculations with the NSC97 potentials 
were concluded \cite{ns97},
which do include extensions to the HH sector,
but yield very similar results.

The presence of hyperons strongly softens the EOS, mainly due to the
larger number of baryonic degrees of freedom. 
This EOS produces a
maximum NS mass that lies slightly below the canonical
value of 1.44 $\ms$ \cite{taylor}, 
which could indicate the presence of non-baryonic (quark) matter 
in the interior of heavy NS \cite{mit,njl,cdm}.
However, the quantitative effects of more reliable NH and HH potentials
as well as of hyperonic TBF need still to be explored in the future.
For these reasons, we will in this article
%present only zero-temperature results with interacting hyperons and 
perform finite-temperature calculations in a much simpler way
by using non-interacting hyperons.
%in order to estimate the sensitivity of the main global observables
%to finite temperature.
This approximation 
%has practically no consequence 
is well justified
for the present
work, because the hadron-quark phase transition occurs at very 
low density, where hyperons do not yet play a role.

In any case, our microscopic baryonic EOS turns out to be very soft and 
features very different characteristics
in comparison to the often used relativistic mean field (RMF) models.
In particular, the combination of a QM phase with a RMF model
leads usually to a reduction of the maximum NS mass \cite{gle,mit}, 
whereas in our approach the maximum mass increases due to the presence of QM,
which is in fact required in order to cover the current observational
values of NS masses.
Also for PNS very different properties will arise,
as we will see.

%------------------------------------------------------------------------------
\subsection{Composition and EOS of hot stellar matter}

For the determination of the composition of beta-stable baryonic matter, 
one needs to know the functional dependence of the free energy,
Eq.~(\ref{e:fb}),
on the individual partial densities and on temperature.
In Ref.~\cite{proto} we have provided analytical parametrizations
of our numerical results for this purpose.
From the free energy density one can then calculate 
the nucleon chemical potentials
\be
 \mu_i = \frac{\partial f}{\partial \rho_i} \:,
\ee
whereas
the chemical potentials of the noninteracting leptons and hyperons
are obtained by solving numerically the free
Fermi gas model at finite temperature. 

For stars in which the strongly interacting particles are only baryons, 
the composition at given baryon density $\rho=\sum_i \rho_i$
is determined by the requirements of beta equilibrium
and charge neutrality, i.e.,
\bea
 \mu_i &=& b_i \mu_n - q_i( \mu_l - \mu_{\nu_l}) \:,
\label{e:mutrap}
\\
 0 &=& \sum_i q_i x_i + \sum_l q_l x_l \:,
%\\
% \rho &=& \sum_i \rho_i \:,
\eea
where $x_i=\rho_i/\rho$ is the baryon concentration,
$b_i$ the baryon number, and $q_i$ the electric charge
of the species $i$.
Equivalent quantities are defined for the leptons $l=e,\mu$.
The initial PNS contains trapped neutrinos, so the electron and muon 
lepton numbers
\be
 Y_l = x_l - x_{\bar l} + x_{\nu_l} - x_{\bar{\nu}_l}
\label{lepfrac:eps}
\ee
are conserved on dynamical time scales
and we fix them to  
$Y_e = 0.4$ and $Y_\mu = 0$, 
as indicated by gravitational collapse calculations of the 
iron core of massive stars.% \cite{dwarf}.

\begin{figure}[t] %............................................................
\includegraphics[width=9cm,angle=270]{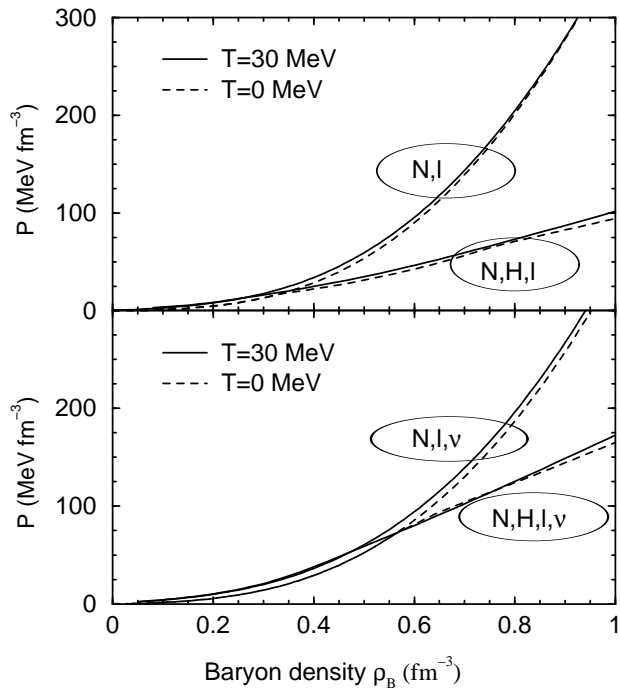}
\caption{
The EOS in beta-stable baryon matter is displayed for temperatures 
$T=0$ and 30 MeV. 
The upper (lower) panel displays results for neutrino-free 
(neutrino-trapped) matter,
for purely nucleonic (upper curves) 
and hyperonic (lower curves) stellar matter.}
\label{f:eos}
\end{figure} %.................................................................

Solving these equations,
we have found in \cite{proto} that the electron fraction is larger in
neutrino-trapped than in neutrino-free matter, and, as a
consequence, the proton population is larger. 
Moreover,
neutrino trapping shifts the threshold density of the $\Sigma^-$ to high
density, whereas $\Lambda$'s appear at slightly smaller density.
This is due to the fact that the $\Sigma^-$ onset depends on the
neutron and lepton chemical potentials, i.e., 
$\mu_{\Sigma^-} = \mu_n + \mu_e - \mu_{\nu_e}$, 
which stays at larger values in neutrino-trapped
matter than in the neutrino-free case
because of the larger fraction of electrons,
thus delaying the appearance
of the $\Sigma^-$ to higher baryon density and limiting its
population to a few percent. 
On the other hand, the $\Lambda$ onset
depends on the neutron chemical potential only, 
$\mu_{\Lambda} = \mu_n$, 
which stays at slightly lower values in the neutrino-trapped case. 
When the temperature increases,
the hyperon thresholds disappear and
more and more hyperons are present also at low densities, but they
represent only a small fraction of the total baryon density in this
region of the PNS. 
Altogether, the hyperon fractions
are much smaller than in the neutrino-free matter 
and therefore the
corresponding EOS will be stiffer than in the neutrino-free case.

Once the composition of the beta-stable stellar matter is known,
one can proceed to calculate the free energy density $f$ and then the
pressure $p$ through the usual thermodynamical relation
\be
 p = \rho^2 \frac{\partial{(f/\rho)}}{ \partial{\rho}} \:.
\ee
The resulting EOS is displayed in Fig.~\ref{f:eos}, where the
pressure for beta-stable stellar matter, without (upper
panel) and with (lower panel) neutrinos, is plotted as a
function of the baryon density at temperatures $T=0$ and 30 MeV.
We notice
that thermal effects produce a slightly stiffer EOS with respect to
the cold case, and that at very high densities they almost play no
role. The inclusion of hyperons, however, causes a dramatic effect, because
the EOS gets much softer, no matter the value of the temperature.
In the neutrino-trapped case, the EOS is slightly softer than in  
neutrino-free matter if
only nucleons and leptons are present in the stellar matter. 
Also here
the presence of hyperons introduces a strong softening of the EOS,
but less than in the neutrino-free case, because now the hyperons
appear later in the matter and their concentration is lower. 
This fact could lead to metastable stars which suffer a delayed collapse 
while cooling down, as discussed in \cite{praka}.

%==============================================================================
\section{Quark matter equation of state}
\label{s:qm}

%------------------------------------------------------------------------------
\subsection{The MIT Bag Model}

We review briefly the description of the bulk properties of
uniform QM at finite temperature, deconfined from the
beta-stable hadronic matter discussed in the previous section, by
using the MIT bag model \cite{chodos}.
In its simplest form, the
quarks are considered to be free inside a bag and the thermodynamic
properties are derived from the Fermi gas model,
where the quark $q=u,d,s$ baryon density, 
the energy density,
and the free energy density 
are given by
\bea
 \rho_q &=& \frac{g}{3} \int\!\!\frac{d^3k}{(2\pi)^3} 
 \left[ f^+_q(k)-f^-_q(k) \right] \:,
\label{e:rhoq}
\\
 \eps_Q &=& g \sum_q \int\!\!\frac{d^3k}{(2\pi)^3}
 \left[ f^+_q(k)+f^-_q(k) \right] E_q(k) + B \:,
\label{e:epsq}
\\
% p &=&  {g\over 3} \sum_q \int\!\!{d^3p\over(2\pi)^3}
% \left[ f^+_q(k) + f^-_q(k) \right] {k^2 \over E_q(k)} - B \:,
%\label{e:pq}
%\\ 
% \omega &=& g T \sum_q \int\!\!{d^3k\over(2\pi)^3}
% \left[ \ln(1-f^+_q(k)) + \ln(1-f^-_q(k)) \right] + B \:,
%\label{e:om} 
 f_Q &=& \eps_Q - T\sum_q s_q \:,      % =\omega + \rho \mu 
\label{e:fq}
%\\
% s_q &=& g \int\!\!{d^3k\over(2\pi)^3}
\eea
where $g=6$ is the quark degeneracy,
$E_q(k) = \sqrt{m_q^2+k^2}$, 
%$m_q$ are the quark masses, 
$B$ is the bag constant,
$s_q$ the entropy density of a noninteracting quark gas,
and the Fermi distribution functions for the quarks and
anti-quarks are
\be
 f^\pm_q(k) = \frac{1}{ 1+\exp[(E_q(k) \mp \mu_q)/T] }
\label{distf}
\ee
with $\mu_q$ being the quark chemical potentials.
The corresponding
expressions at $T=0$ can be obtained by eliminating the antiparticles and
substituting the particle distribution functions by the usual step
functions. 
We have used massless $u$ and $d$ quarks, and $m_s=150$ MeV.

It has been found \cite{alford,mit} that within the MIT bag model
(without color superconductivity) with a density-independent bag
constant $B$, the maximum mass of a NS cannot exceed a value of
about 1.6 solar masses. 
Indeed, the maximum mass increases as the
value of $B$ decreases, but too small values of $B$ are incompatible
with a hadron-quark transition density $\rho >$ 2--3 $\rho_0$ in 
nearly symmetric nuclear matter, 
as demanded by heavy-ion collision phenomenology. 
Values of $B\gtrsim 150\;\rm MeV\!/fm^{-3}$ 
can also be excluded within our model, 
since we do not obtain any more a phase transition in beta-stable matter 
in combination with our baryonic EOS \cite{mit}.

In order to overcome these restrictions of the model, 
one can introduce a density-dependent
bag parameter $B(\rho)$, and this approach was followed in
Ref.~\cite{mit}. 
This allows one to lower the value of $B$ at large
density, providing a stiffer QM EOS and increasing the
value of the maximum mass, while at the same time still fulfilling
the condition of no phase transition below $\rho \approx 3 \rho_0$
in symmetric matter.
In the following we present results based on the MIT model using both 
a constant value of the bag parameter, $\bc$, and a
gaussian parametrization for the density dependence,
\be
 {B(\rho)} = B_\infty + (B_0 - B_\infty)
 \exp\left[-\beta\Big(\frac{\rho}{\rho_0}\Big)^2 \right]
\label{eq:param} \ee 
with $B_\infty = 50\;\rm MeV\!/fm^{-3}$, $B_0 =
400\;\rm MeV\!/fm^{3}$, and $\beta=0.17$,
see Ref.~\cite{mit}.

The introduction of a density-dependent bag 
has to be taken into account properly for the 
computation of various thermodynamical quantities; 
in particular the quark chemical potentials
and the pressure are modified as
\bea
 \mu_q &\rightarrow& \mu_q + \frac{dB(\rho)}{ d\rho} \:,
\label{murho}
\\
 p &\rightarrow& p + \rho \frac{dB(\rho)}{d\rho} \:.
\eea
Nevertheless, due to a cancelation of the second term in (\ref{murho}), 
occurring in relations (\ref{eq:chem}) for the beta-equilibrium, 
the composition at a given total baryon density remains unaffected
by this term (and is in fact independent of $B$).
At this stage of investigation, we disregard possible dependencies 
of the bag parameter on temperature and individual quark densities.
For a more extensive discussion of this topic, the reader is referred to 
Refs.~\cite{mit,njl,cdm}.

%------------------------------------------------------------------------------
\subsection{Quark matter in beta equilibrium}

In a PNS with QM and trapped neutrinos we must 
add the contribution of the
leptons as free Fermi gases to the (free) energy density, 
Eqs.~(\ref{e:epsq},\ref{e:fq}),
and impose beta equilibrium, charge neutrality,
and baryon and lepton number conservation \cite{gle}.

More precisely, the individual quark chemical potentials
are fixed by Eq.~(\ref{e:mutrap}) with $b_q=1/3$, 
which implies
\be
 \mu_d = \mu_s = \mu_u + \mu_l - \mu_{\nu_l} \:.
\label{eq:chem} 
\ee
The charge neutrality condition and the total
baryon number conservation read
\bea
 0 &=& 2\rho_u - \rho_d - \rho_s - \rho_e \:,
\\
 \rho &=& \rho_u + \rho_d + \rho_s \:,
\label{eq:baryon} 
\eea
and Eq.~(\ref{lepfrac:eps}) specifies lepton number
conservation.
These equations determine the composition $\rho_q(\rho)$
and the pressure of the QM phase, 
\be
 p_Q(\rho) = \rho \frac{\partial f_Q}{ \partial\rho} - f_Q \:.
\ee

\begin{figure}[t] %............................................................
\includegraphics[width=8cm,clip]{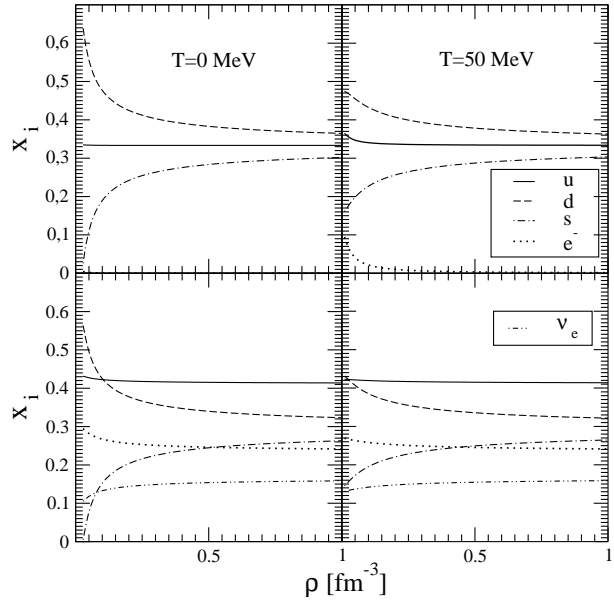}
\caption{ 
Particle fractions in beta-stable QM at $T=0$
(left panels) and $T=50\;\rm MeV$ (right panels) for neutrino-free
(upper panels) and neutrino-trapped (lower panels) matter.}
\label{f:xi}
\end{figure} %.................................................................

\begin{figure*}[t]
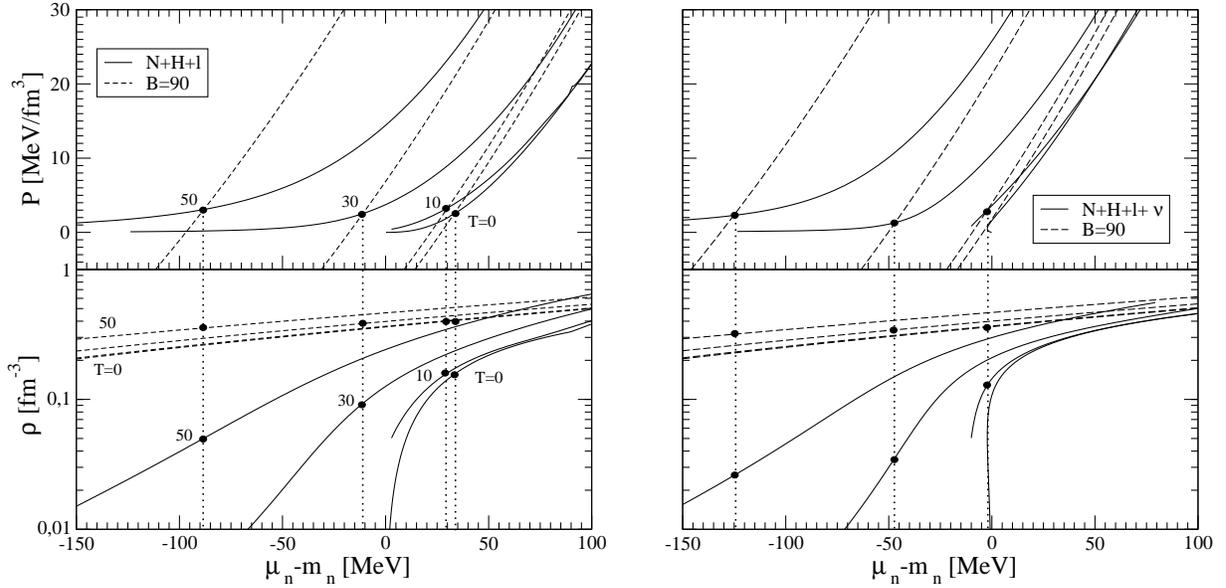
 %...........................................................
\centering
\subfigure{\includegraphics[width=8cm,clip]{figure3a.eps}}
\hspace{-1mm}
\subfigure{\includegraphics[width=8cm,clip]{figure3b.eps}}
\caption{ 
Baryon density (lower panels) and pressure (upper panels)
as a function of baryon chemical potential of beta-stable baryonic
matter (solid curves) and quark matter (dashed curves) with (right
panels) and without (left panels) neutrino trapping at different
temperatures $T=0,10,30,50\;\rm MeV$. The vertical dotted lines
indicate the positions of the phase transitions. A bag constant
$\bc$ is used for QM.} 
\label{f:pc}
\end{figure*} %................................................................

\begin{figure*}[t] %...........................................................
\centering
\subfigure{\includegraphics[width=8cm,clip]{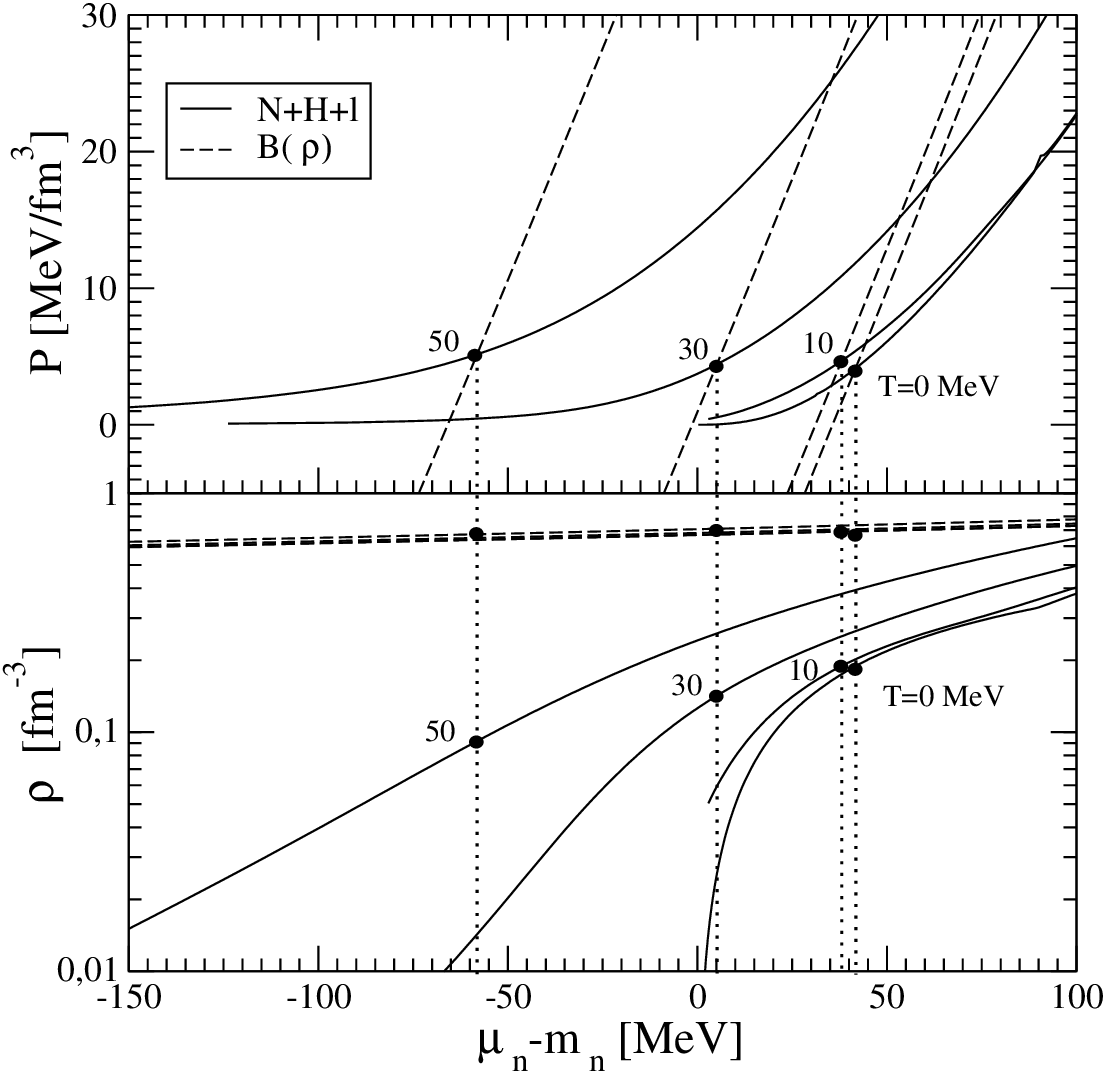}} 
\hspace{-1mm}
\subfigure{\includegraphics[width=8cm,clip]{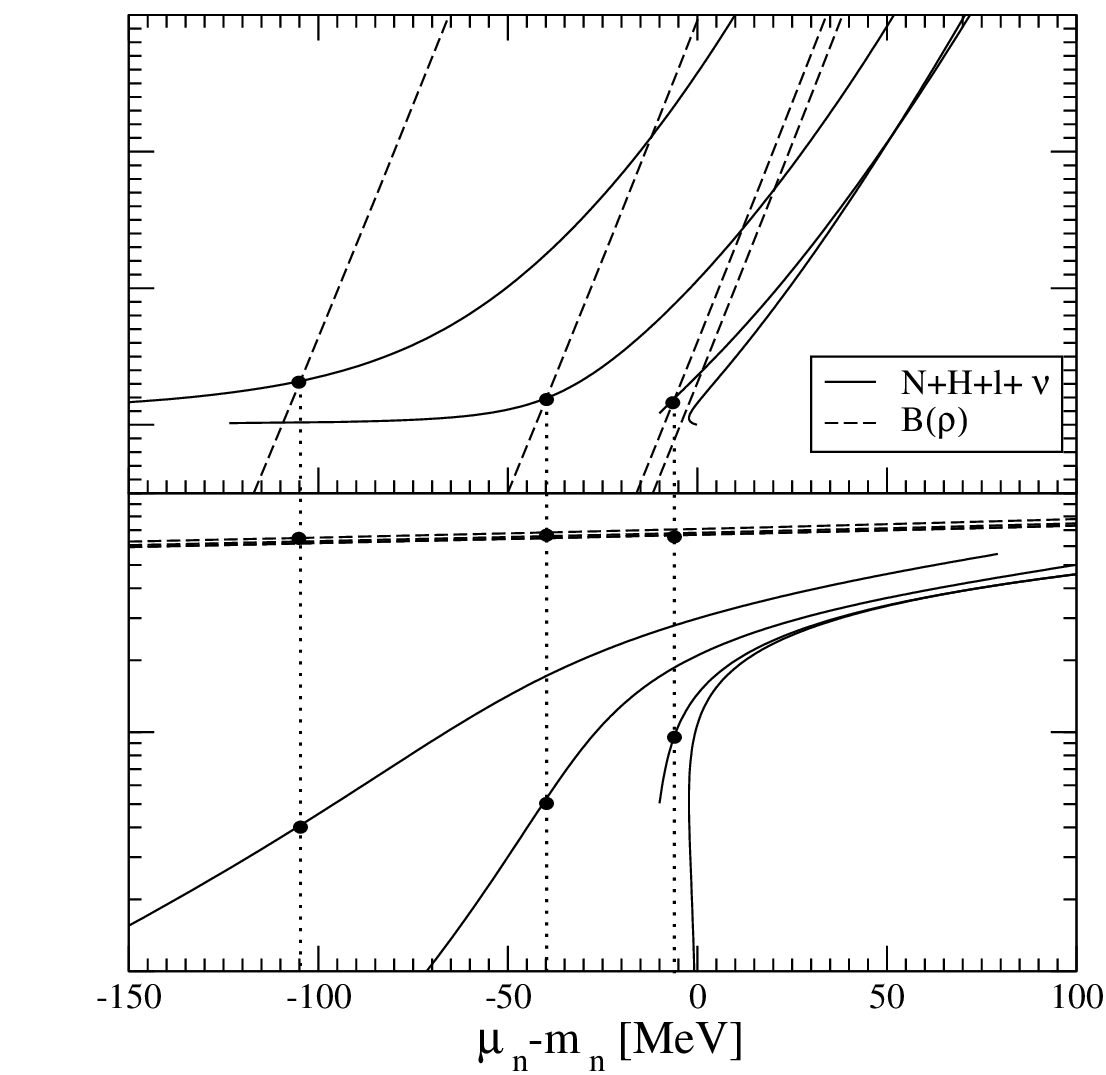}}
\caption{ 
Same as Fig.~\ref{f:pc}, but with a density-dependent bag parameter.} 
\label{f:pr}
\end{figure*} %................................................................

\begin{figure}[t] %............................................................
\includegraphics[width=8cm]{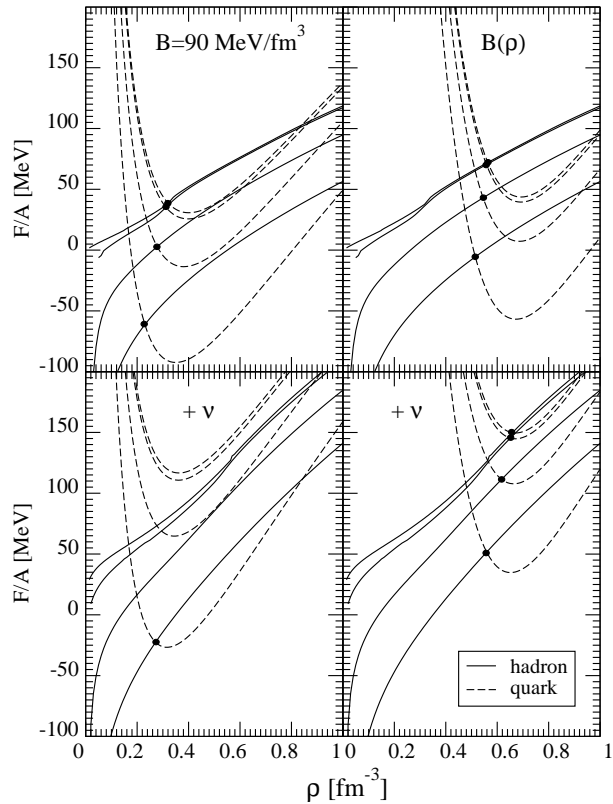}
\caption{ 
Free energy per baryon as a function of baryon density of beta-stable
baryonic matter (solid lines) and quark matter (dashed lines) with a
bag constant $\bc$ (left panels) or a density-dependent bag
parameter (right panels) with (lower panels) and without (upper
panels) neutrino trapping at different temperatures
$T=0,10,30,50\;\rm MeV$ (upper to lower curves).} 
\label{f:f}
\end{figure} %.................................................................

\begin{figure*}[t] %...........................................................
\includegraphics[width=12cm,clip]{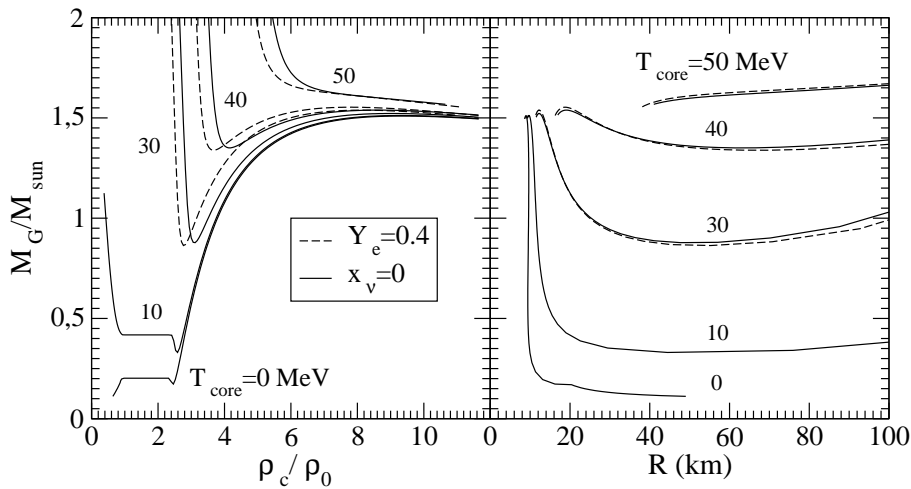}
\caption{ 
(Proto)Neutron star mass-central density (left panel) and
mass-radius (right panel) relations for different core temperatures
$T=0,10,30,40,50\;\rm MeV$ and neutrino-free (solid curves) or
neutrino-trapped (dashed curves) matter. A bag constant $\bc$ is
used for QM.} 
\label{f:mc}
\end{figure*} %................................................................

\begin{figure*}[t] %...........................................................
\includegraphics[width=12cm,clip]{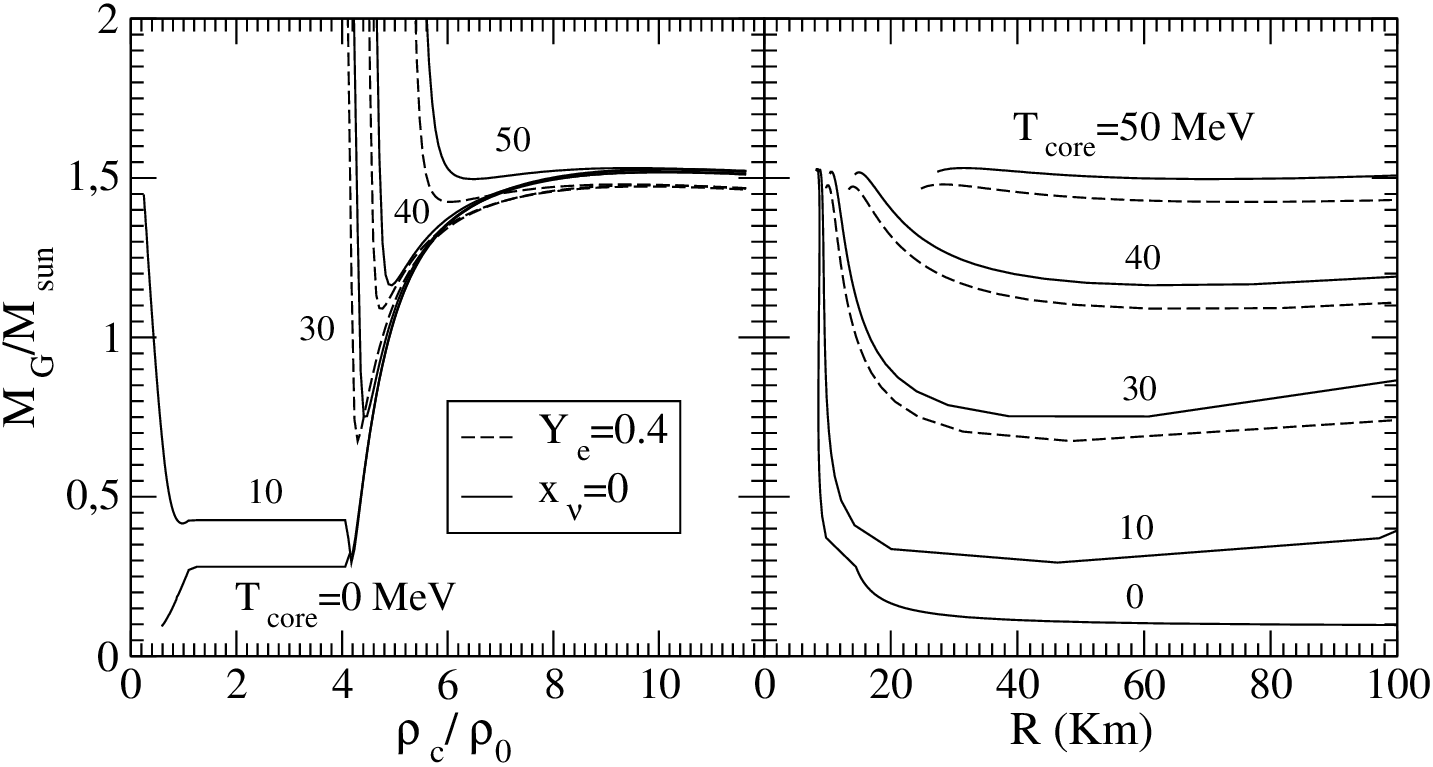}
\caption{ 
Same as Fig.~\ref{f:mc}, but with a density-dependent bag
parameter.} 
\label{f:mr}
\end{figure*} %................................................................

\begin{figure}[t] %............................................................
\includegraphics[width=8cm,clip]{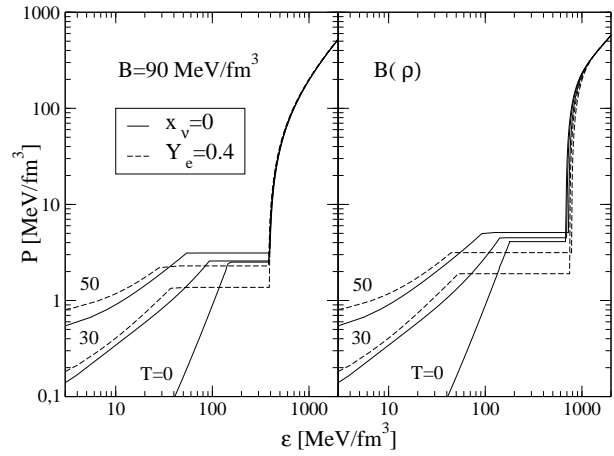}
\caption{ 
Pressure as a function of energy density for beta-stable matter
with (dashed curves) and without (solid curves) neutrino trapping 
at different temperatures $T=0$, 30, and 50 MeV
with a bag constant $\bc$ (left panel) 
or a density-dependent bag parameter (right panel).}
\label{f:pe}
\end{figure} %.................................................................

\begin{figure}[htbp]
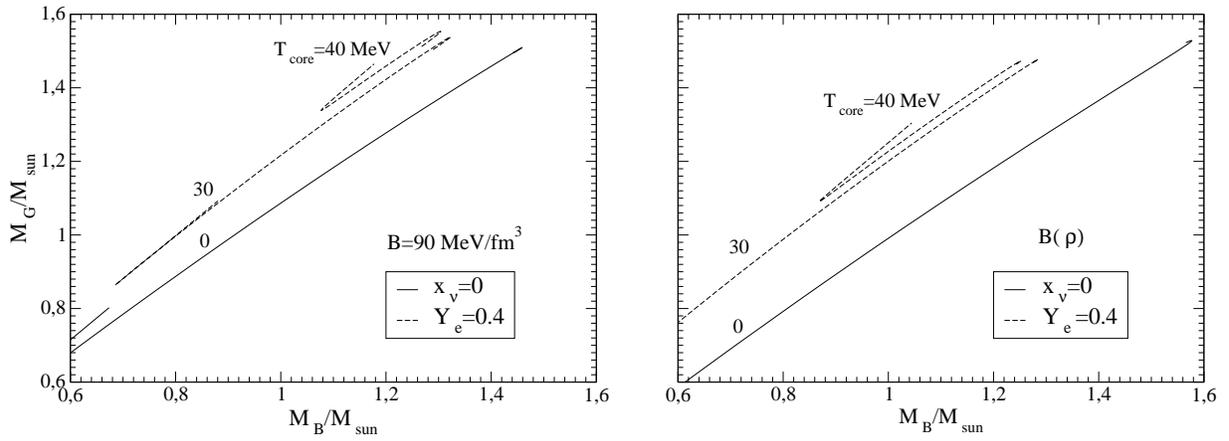
 %............................................................
\subfigure{\includegraphics[width=8.0cm,clip]{figure9a.eps}}
%\vspace{-10mm}
\subfigure{\includegraphics[width=8.0cm,clip]{figure9b.eps}} %8.15
\caption{ 
Relation between gravitational mass and baryonic mass for
cold neutron stars (solid curves)
and protoneutron stars at $T=30,40\;\rm MeV$ (dashed lines)
with a bag constant $\bc$ (upper panel) 
or a density-dependent bag parameter (lower panel).}
\label{f:mb}
\end{figure} %.................................................................

In Fig.~\ref{f:xi} we plot the particle fractions
$x_i=\rho_i/\rho$ as a function of baryon density for
neutrino-free (upper panels) and neutrino-trapped (lower panels) QM. 
Since in the range of temperature considered here
thermal effects are rather weak, we report only results for $T=0$
and $T=50$ MeV. 
Conversely, the presence of neutrinos influences quite strongly
the composition:
In this case the relative
fraction of $u$ quarks increases substantially from $33\%$ to about
$42\%$, compensating the charge of the electrons that are present at
an average percentage of $25\%$ throughout the considered range of
baryon density, whereas $d$ and $s$ quark fractions are slightly lowered. 
One also can argue from the figure that most of
the electrons present in the neutrino-free case come from thermally
excited pairs.

Before studying the phase transition inside PNS, we
would like to reiterate the fact that our combination of baryonic
and QM EOS does not allow any phase transition in
symmetric nuclear matter below about 3--4 $\rho_0$, as ruled
out by observational evidence. This has been amply discussed in
Refs.~\cite{mit,njl,cdm} and was actually an important motivation
for our choice of the QM EOS,
as mentioned above.

%------------------------------------------------------------------------------
\subsection{Phase transition in hot beta-stable matter}

We now consider the hadron-quark phase transition in 
beta-stable matter at finite temperature.
In the present work we adopt the
simple Maxwell construction for the phase transition from the plot
of pressure versus chemical potential. 
The more general Glendenning (Gibbs) construction \cite{gle,glen} 
is still affected by many theoretical uncertainties \cite{mixed}
and in any case influences very little the final
mass-radius relations of massive (proto)neutron stars \cite{mit}.

We therefore display in Figs.~\ref{f:pc} and \ref{f:pr} the pressure
$p$ (upper panels) and baryon density $\rho$ (lower panels) as
functions of the baryon chemical potential $\mu_n$ for both
baryonic and QM phases at temperatures $T=0,10,30,50$ MeV.
The crossing points of the baryon and quark pressure curves (marked
with a dot) represent the transitions between baryon and QM
phases. The projections of these points (dotted lines) on the baryon
and quark density curves in the lower panels indicate the
corresponding transition densities from low-density baryonic matter,
$\rho_H$, to high-density QM, $\rho_Q$. The results in
Fig.~\ref{f:pc} are obtained with $\bc$ and those in Fig.~\ref{f:pr}
with $B(\rho)$.

The following general observations can be made:\\
(i) The transition density $\rho_H$ is rather low,
of the order of the nuclear matter saturation density;\\
(ii) The phase transition density jump $\rho_Q-\rho_H$ is large,
several times $\rho_H$;\\
(iii) Thermal effects shift $\rho_H$ to lower values
of subnuclear densities and increase the density jump $\rho_Q-\rho_H$;\\
(iv) Neutrino trapping lowers even more the transition density. 
For the cold case the presence of neutrinos even inhibits completely the
phase transition. 
However, this feature has no physical
relevance, because no trapping occurs in cold catalyzed neutron stars.
The reason for such behavior is the stronger increase of the QM pressure 
due to neutrino trapping compared to the one of the baryonic phase,
as evidenced in the figure.
This leads naturally to an earlier onset of the QM phase;\\
(v) The model with density-dependent bag parameter predicts 
(by construction, see the discussion above)
larger transition densities $\rho_H$ and larger jumps 
$\rho_Q-\rho_H$ than those with bag constant $\bc$. 
The plateaus in the Maxwell
construction are thus wider for the former case.

These qualitative conclusions are reaffirmed in Fig.~\ref{f:f},
where we report the free energy per baryon, $F/A$, as a function of
baryon density, the four panels corresponding to the cases mentioned
above with and without neutrino trapping and with different bag
parameters. Here the crossing points between solid lines (baryon
phase) and dashed lines (QM phase) indicate an average
phase transition density, above which the QM phase is
characterized by a value of the free energy lower than that of the
baryon phase, thus indicating that the QM phase is energetically
favoured.

Based on these results for the beta-stable baryon and QM phases, 
we proceed now to the determination of the properties of
static (proto)neutron stars.

%==============================================================================
\section{Protoneutron star structure}
\label{s:res}
We assume that a (proto)neutron star is a spherically symmetric
distribution of mass in hydrostatic equilibrium. The equilibrium
configurations are obtained by solving the
Tolman-Oppenheimer-Volkoff (TOV) equations \cite{shapiro} for the
pressure $p$ and the enclosed gravitational mass $m$ 
and baryonic mass $m_B$, 
\bea
  \frac{dp(r)}{ dr} &=& -\frac{ G m(r) \epsilon(r)}{ r^2 }
\nonumber\\ && \times
  { \left[ 1 + {p(r) / \epsilon(r)} \right]
  \left[ 1 + {4\pi r^3 p(r) / m(r)} \right]
  \over
  1 - {2G m(r)/ r} } \:,\quad
\\
 {dm(r) \over dr} &=& 4 \pi r^2 \epsilon(r) \:,
\\
 {dm_B(r)\over dr} &=& \frac{4\pi r^2 \rho(r)\, m_N}{\sqrt{1-2Gm(r)/r}} \:,
\eea 
where $m_N= 1.67\times 10^{-24}$g is the nucleon mass and 
$G$ the gravitational constant. Starting with a central
mass density $\epsilon(r=0) \equiv \epsilon_c$, we integrate out
until the pressure on the surface equals the one corresponding to
the density of iron. This gives the stellar radius $R$ and the
gravitational mass $M_G \equiv m(R)$ and
baryonic mass $M_B \equiv m_B(R)$.

Moreover, in our model we assume that a PNS, in its early stage, is composed 
of a hot
isothermal and neutrino-opaque core separated from an outer cold crust 
(described by the Baym-Pethick-Sutherland \cite{bps}  
and the Feynman-Metropolis-Teller \cite{fmt} EOS) 
by an isentropic,
beta-equilibrated, and neutrino-free intermediate layer in the range
of baryon density from $0.01\;\rm fm^{-3}$ down to $10^{-6}\;\rm
fm^{-3}$ \cite{nicot1}, which is based on the EOS LS220 of Lattimer and Swesty
\cite{lat} with compressibility $K=220$ MeV. For the isothermal core
we use as input the BHF EOS for the baryonic matter and the MIT bag
models for the beta-stable QM phase, as discussed above. In order
to ensure an exact matching of all the thermodynamic quantities
between core and crust we perform a fine tuning of the entropy per baryon
of the isentropic envelope, $S_{\rm env.}$, around the matching value of
baryon density $\rho_{\rm env.}\approx 0.01\;\rm fm^{-3}$. More precisely,
fixing $S_{\rm env.}$ to the values $6,8,10$ (in units of the
Boltzmann constant) we get temperature profiles which drop quickly
from $T_{\rm core}=30,40,50$ MeV to zero in the considered range of
density, thus suggesting a natural correspondence between values of
$T_{\rm core}$ and $S_{\rm env.}$ \cite{nicot1}.

In other words, we model the PNS by adopting a hybrid
isothermal plus isentropic temperature profile, allowing us to take into
account as much as possible information coming from dynamical
simulations \cite{praka,burr,strobel,pons}, 
which show that the temperature
drops rapidly to zero at the surface of the star due to the fast
cooling of the outer part of the PNS, where the stellar matter is
transparent to neutrinos. In addition, during the early stage, the
outer part of a PNS is characterized by a high value of the entropy
per baryon. Essentially, in the first ten seconds, the entropy
profile decreases from the surface to the core starting from values
of 6--10 \cite{burr}.

We consider this approach more realistic than the one employed for
baryonic stars in \cite{proto}, where a temperature profile in the
shape of a step function was assumed, joining directly the baryonic
EOS and the one for the cold outer crust. This procedure leads,
however, to unstable PNS at $T_{\rm core}$ slightly above $30$ MeV
and also yields a strong dependence on the value of density where
the cold crust is attached of the minimum PNS mass, which is a
fundamental quantity characterizing the range of stability of such
hot and compact objects. The improved procedure described above
applied to baryonic PNS leads to more reliable values of the minimum
gravitational mass, while hardly affecting the value of the maximum
mass, which mainly depends on the EOS employed and the core
composition. For the purely nucleonic case (no hyperons) we find now a
maximum mass within a narrow range of 1.77--1.87 $\ms$ \cite{nicot1},
similar to \cite{proto}.

In the following we present results schematizing the entire
evolution of the star as divided in two main stages. 
The first consists of a PNS with a hot 
($T_{\rm core}\approx$ 30--50MeV) 
neutrino-trapped core and a high-entropy transition layer
($S_{\rm env.}\approx$ 6--10),
joined to a cold outer crust.
The second stage represents the short-term cooling, 
where the neutrino-free core possesses a low temperature of about $10$ MeV 
%and the inner crust can be considered cold ($S_{\rm env.}\approx 0$). 
and is direcly attached to a cold crust at 
$\rho = 3\times10^{-4}\;\rm fm^{-3}$.
%$\rho = 0.0003\;\rm fm^{-3}$.

%------------------------------------------------------------------------------
\subsection{Numerical Results}

The results are plotted in Figs.~\ref{f:mc} and \ref{f:mr}, where we
display the gravitational mass $M_G$ (in units of the solar mass
$\ms$) as a function of the radius $R$ (right panels) and the central
baryon density $\rho_c$ (left panels), for QM EOS with
$\bc$ and $B(\rho)$, respectively.

Due to the use of the Maxwell construction, the curves are not
continuous \cite{gle}: For small enough central densities (large
radii) the stars are purely baryonic. Then a sudden increase of the
central density is required in order to initiate the QM phase in the
center of the star, corresponding to the phase diagrams
Figs.~\ref{f:pc} and \ref{f:pr}.
%Due to this effect, also a discontinuity in the possible radii arises.
By performing the Glendenning construction, the curves would become
continuous. Heavy PNS in our approach are thus practically quark
stars with only a thin outer layer of baryonic matter.
Below the maximum mass configuration, however, the stars develop
an extended outer envelope of hot matter, 
the details of which depend on the treatment of the low-density 
baryonic phase and the phase transition.
We will therefore in the following focus on the properties of
heavy stars close to the limiting mass,
which are unaffected by these complications.

For completeness we display the complete set of results at core
temperatures $T=0,10,30,40,50\;\rm MeV$ with and without neutrino
trapping, although only the curves with high temperatures and
neutrino trapping and low temperatures without trapping are the
physically relevant ones. We observe in any case a surprising
insensitivity of the results to the presence of neutrinos, in 
particular for the $\bc$ case, which can
be traced back to the fact that the QM EOS $p(\eps)$ 
is practically insensitive to the neutrino fraction.
In fact, for the $\bc$ case and assuming massless quarks and leptons,
the universal relation $p=(\eps-4B)/3$ 
would hold irrespective of the internal composition 
of the quark-lepton phase.
This is illustrated in Fig.~\ref{f:pe}, 
showing the beta-stable EOS $p(\eps)$ at three temperatures
with and without neutrinos.
Indeed the pressure in the QM phase is nearly the same 
in both cases,even to a lesser degree for the $B(\rho)$ EOS.

On the other hand, the temperature dependence of the curves is quite
pronounced for intermediate and low-mass stars, showing a strong
increase of the minimum mass with temperature, whereas the maximum
mass remains practically constant under all possible circumstances.
Above core temperatures of about 40--50 MeV all stellar configurations
become unstable.

Concerning the dependence on the QM EOS, we observe again
only a slight variation of the maximum PNS masses between $1.55\;\ms$
for $\bc$ and $1.48\;\ms$ for $B(\rho)$. Clearer differences exist for
the radii, which for the same mass and temperature are larger for
the $\bc$ model, as has also been found in \cite{mit,cdm} for cold NS.
The maximum PNS masses turn out to be very close to those
of cold NS, thus excluding the possibility of metastable configurations.

In order to illustrate better this issue,
we show in Fig.~\ref{f:mb} the relation between baryonic and 
gravitational mass for cold NS and PNS at $T=30$ and 40 MeV.
In this plot
the evolution of an isolated star proceeds on vertical lines connecting
the upper curve for PNS with the lower curve for NS.
One notes immediately that metastable hybrid stars do not exist
in our model:
For the $B(\rho)$ case,
the heaviest PNS ($M_B=1.28\ms,\ M_G=1.48\ms$) 
transits into a $M_G=1.26\ms$ NS,
which is consequently the heaviest NS that can be produced 
without accretion in our approach.
Accretion might then further augment the NS mass to a 
maximum stable value of $M_G=1.53\ms$.

%==============================================================================
\section{Conclusions}
\label{s:end}

In this article we extended our previous works on cold baryonic
\cite{hypns} and hybrid \cite{mit,njl,cdm} NS and
baryonic PNS \cite{proto} to the case of hybrid PNS.
We combined the most recent microscopic baryonic EOS in
the BHF approach involving nuclear three-body forces and hyperons
with two versions of a generalized MIT bag model describing the
QM phase: One using a fixed bag constant, the other one a
density-dependent bag parameter $B(\rho)$ in order to explore the
maximum PNS mass that can be reached in this approach.

We modelled the profile of a PNS in an extremely
simplified way with a constant lepton fraction $Y_e=0.4$ and
constant temperature of the core and an isentropic cover,
leaving the core temperature as a global parameter.
A really satisfying treatment would require coupled dynamical
simulations for the various microscopic and macroscopic 
evolution equations,
which is currently beyond our reach.

We found in Ref.~\cite{proto} that purely baryonic (hyperonic) PNS
can reach masses of about $1.5\;\ms$, and nearly the same mass limit
is now obtained also for hybrid PNS. The difference between both
configurations lies in the transition to a cold NS, which for
baryonic stars could lead to a delayed collapse to a black hole, as
the cold baryonic NS cannot support masses up to
$1.5\;\ms$ \cite{hypns}, whereas this phenomenon is exluded for
hybrid NS, which can sustain masses up to those of the PNS.

This result is in contrast to Ref.~\cite{cooke}, where such
metastability was found also for hybrid stars;
however, with a much stiffer baryonic EOS and 
correspondingly larger value of the bag constant $B=200\;\rm MeV\!/fm^3$,
which is excluded in our model.
As mentioned before, our baryonic EOS is especially soft 
due to the presence of hyperons, compensating the
repulsive character of nucleonic TBF at high density.
Its associated maximum NS mass remains below 1.4 solar masses, 
and the presence of QM inside the star is {\em required} 
in order to reach larger maximum masses.
We have thus shown that the presumed increase of the hadron-quark
phase transition density due to the presence of neutrino trapping 
\cite{cooke,lugon,nemes}
is not a general feature, but might be reversed in combination with 
a very soft baryonic EOS.

Altogether, we once again confirm our prediction of rather low
limiting masses for (proto)neutron stars, irrespective of variations
of the baryonic or QM EOS. Therefore, the experimental
observation of a very heavy ($M \gtrsim 1.8 \ms$) NS, as
claimed recently by some groups \cite{kaaret} ($M \approx 2.2 \ms$), 
if confirmed, could hint to serious problems 
for the current rather simple theoretical modelling of the
high-density quark matter phase, 
the assumptions about the phase transition between baryon and QM phase,
or even for the TOV equations at extreme baryon densities.

%%%%%%%%%%%%%%%%%%%%%%%%%%%%%%%%%%%%%%%%%%%%%%%%%%%%%%%%%%%%%%%%%%%%%%%%%%%%%%%


\begin{thebibliography}{99}

\bibitem{shapiro}
 S. L. Shapiro and S. A. Teukolsky,
 {\em Black Holes, White Dwarfs, and Neutron Stars}
 (John Wiley and Sons, New York, 1983).

\bibitem{bethe}
% SUPERNOVA MECHANISMS
 H. A. Bethe, 
 Rev. Mod. Phys. {\bf 62}, 801 (1990).

\bibitem{proto}
% PROTONEUTRON STARS WITHIN THE BRUECKNER-BETHE-GOLDSTONE THEORY
 O. E. Nicotra, M. Baldo, G. F. Burgio, and H. J. Schulze,
 Astron. Astrophys. {\bf 451}, 213 (2006).

\bibitem{thl}
% THREE-BODY CORRELATIONS IN NUCLEAR MATTER
 B. D. Day,
 Phys. Rev. {\bf C24}, 1203 (1981);
% BETHE-BRUECKNER-GOLDSTONE EXPANSION IN NUCLEAR MATTER
 H. Q. Song, M. Baldo, G. Giansiracusa, and U. Lombardo,
 Phys. Rev. Lett. {\bf 81}, 1584 (1998);
% BETHE-BRUECKNER-GOLDSTONE EXPANSION IN NEUTRON MATTER
 M. Baldo, G. Giansiracusa, U. Lombardo, and H. Q. Song,
 Phys. Lett. {\bf B473}, 1 (2000);
% HIGH DENSITY SYMMETRIC NUCLEAR MATTER IN THE BBG APPROACH
 M. Baldo, A. Fiasconaro, H. Q. Song, G. Giansiracusa, and U. Lombardo,
 Phys. Rev. {\bf C65}, 017303 (2002);
% SOLUTION TO THE BETHE-FADDEEV EQUATION WITHIN THE CONTINUOUS VERSION
% OF THE HOLE-LINE EXPANSION
 R. Sartor,
 Phys. Rev. {\bf C73}, 034307 (2006).

\bibitem{hypmat}
% HYPERNUCLEAR MATTER IN THE BRUECKNER-HARTREE-FOCK APPROXIMATION
 H.-J. Schulze, A. Lejeune, J. Cugnon, M. Baldo, and U. Lombardo,
 Phys. Lett. {\bf B355}, 21 (1995);
% HYPERONIC NUCLEAR MATTER IN BRUECKNER THEORY
 H.-J. Schulze, M. Baldo, U. Lombardo, J. Cugnon, and A. Lejeune,
 Phys. Rev. {\bf C57}, 704 (1998).

\bibitem{hypns}
% ONSET OF HYPERON FORMATION IN NEUTRON STAR MATTER FROM BRUECKNER THEORY
 M. Baldo, G. F. Burgio, and H.-J. Schulze,
 Phys. Rev. {\bf C58}, 3688 (1998);
% HYPERON STARS IN THE BRUECKNER-BETHE-GOLDSTONE THEORY
% M. Baldo, G. F. Burgio, and H.-J. Schulze,
 Phys. Rev. {\bf C61}, 055801 (2000),
\bibitem{ns97}
% MAXIMUM MASS OF NEUTRON STARS
 H.-J. Schulze, A. Polls, A. Ramos, and I. Vida\a~na,
 Phys. Rev. {\bf C73}, 058801 (2006).
%% HYPERON-HYPERON INTERACTIONS AND PROPERTIES OF NEUTRON STAR MATTER (???)
% I. Vida\a~na, A. Polls, A. Ramos, L. Engvik, and M. Hjorth-Jensen,
% Phys. Rev. {\bf C62}, 035801 (2000).

\bibitem{fried}
% HOT AND COLD, NUCLEAR AND NEUTRON MATTER
 B. Friedman and V. R. Pandharipande,
 Nucl. Phys. {\bf A361}, 502 (1981).

\bibitem{lej}
% HOT NUCLEAR MATTER IN AN EXTENDED BRUECKNER APPROACH
 A. Lejeune, P. Grang\a'e, M. Martzolff, and J. Cugnon,
 Nucl. Phys. {\bf A453}, 189 (1986).

\bibitem{nic}
% NUCLEAR LIQUID-GAS PHASE TRANSITION
 M. Baldo and L. S. Ferreira,
 Phys. Rev. {\bf C59}, 682 (1999);
% THE LIMITING TEMPERATURE OF HOT NUCLEI FROM MICROSCOPIC EQUATION OF STATE
 M. Baldo, L. S. Ferreira, and O. E. Nicotra,
 Phys. Rev. {\bf C69}, 034321 (2004).

\bibitem{chi}
% CHIRAL DYNAMICS AND NUCLEAR MATTER
 N. Kaiser, S. Fritsch, and W. Weise,
 Nucl. Phys. {\bf A697}, 255 (2002).

\bibitem{malfliet}
 B. Ter Haar and R. Malfliet,
% EOS OF NUCLEAR MATTER IN THE RELATIVISTIC DIRAC-BRUECKNER APPROACH
 Phys. Rev. Lett. {\bf 56}, 1237 (1986);
%
 Phys. Rep. {\bf 149}, 207 (1987).

\bibitem{huber}
% SYMMETRIC AND ASYMMETRIC NUCLEAR MATTER IN THE RELATIVISTIC APPROACH
% AT FINITE TEMPERATURES
 H. Huber, F. Weber, and M. K. Weigel,
 Phys. Rev. {\bf C57}, 3484 (1999).

\bibitem{bloch}
 C. Bloch and C. De Dominicis,
% Un développement du potentiel de gibbs d'un système quantique composé
% d'un grand nombre de particules
 Nucl. Phys., {\bf 7}, 459 (1958);
% ... II, III-La contribution des collisions binaires
 {\bf 10}, 181,509 (1959).

\bibitem{v18}
% ACCURATE NUCLEON-NUCLEON POTENTIAL WITH CHARGE-INDEPENDENCE BREAKING
 R. B. Wiringa, V. G. J. Stoks, and R. Schiavilla,
 Phys. Rev. {\bf C51}, 38 (1995).

\bibitem{book}
% THE MANY-BODY THEORY OF THE NUCLEAR EQUATION OF STATE 
 M. Baldo, 
 {\em Nuclear Methods and the Nuclear Equation of State},
 (World Scientific, Singapore, 1999),
 International Review of Nuclear Physics, Vol. 8.

\bibitem{myers}
% NUCLEAR PROPERTIES ACCORDING TO THE THOMAS-FERMI MODEL
 W. D. Myers and W. J. Swiatecki,
 Nucl. Phys. {\bf A601}, 141 (1996);
% NUCLEAR EQUATION OF STATE
 Phys. Rev. {\bf C57}, 3020 (1998).

\bibitem{uix}
% THREE-NUCLEON INTERACTION IN 3-, 4- AND INFINITE-BODY SYSTEMS
 J. Carlson, V. R. Pandharipande, and R. B. Wiringa,
 Nucl. Phys. {\bf A401}, 59 (1983);
% MOMENTUM DISTRIBUTIONS IN A = 3 AND 4 NUCLEI
 R. Schiavilla, V. R. Pandharipande, and R. B. Wiringa,
 Nucl. Phys. {\bf A449}, 219 (1986).

\bibitem{bbb}
% MICROSCOPIC NUCLEAR EQUATION OF STATE WITH THREE-BODY FORCES AND
% NEUTRON STAR STRUCTURE
 M. Baldo, I. Bombaci, and G. F. Burgio,
 Astron. Astrophys. {\bf 328}, 274 (1997);
% THREE-BODY FORCES AND NEUTRON STAR STRUCTURE
 X. R. Zhou, G. F. Burgio, U. Lombardo, H.-J. Schulze, and W. Zuo,
 Phys. Rev. {\bf C69}, 018801 (2004).

\bibitem{glenhyp}
 N. K. Glendenning,
% THE HYPERON COMPOSITION OF NEUTRON STARS
 Phys. Lett. {\bf B114}, 391 (1982);
% NEUTRON STARS ARE GIANT HYPERNUCLEI?
 Astrophys. J. {\bf 293}, 470 (1985).
%% VACUUM POLARIZATION EFFECTS ON NUCLEAR MATTER AND NEUTRON STARS
% Nucl. Phys. {\bf A493}, 521 (1989);

\bibitem{gle}
 N. K. Glendenning,
 {\em Compact Stars, Nuclear Physics, Particle Physics, and General Relativity},
 2nd ed., 2000, Springer-Verlag, New York.

\bibitem{maessen}
% SOFT-CORE BARYON-BARYON ONE-BOSON-EXCHANGE MODELS.
% II. HYPERON-NUCLEON POTENTIAL
 P. M. M. Maessen, Th. A. Rijken, and J. J. de Swart,
 Phys. Rev. {\bf C40}, 2226 (1989).

\bibitem{taylor}
% DISCOVERY OF A PULSAR IN A BINARY SYSTEM
 R. A. Hulse and J. H. Taylor,
 Astrophys. J. {\bf 195}, L51 (1975);
% FURTHER EXPERIMENTAL TESTS OF RELATIVISTIC GRAVITY USING THE
% BINARY PULSAR PSR 1913+16
 J. H. Taylor and J. M. Weisberg,
 Astrophys. J. {\bf 345}, 434 (1989).

\bibitem{mit}
% MAXIMUM MASS OF NEUTRON STARS WITH A QUARK CORE
 G. F. Burgio, M. Baldo, P. K. Sahu, A. B. Santra, and H.-J. Schulze,
 Phys. Lett. {\bf B526}, 19 (2002);
% HADRON-QUARK PHASE TRANSITION IN DENSE MATTER AND NEUTRON STARS
 G. F. Burgio, M. Baldo, P. K. Sahu, and H.-J. Schulze,
 Phys. Rev. {\bf C66}, 025802 (2002).
\bibitem{njl}
% NEUTRON STARS AND THE TRANSITION TO COLOR SUPERCONDUCTING QUARK MATTER
 M. Baldo, M. Buballa, G. F. Burgio, F. Neumann, M. Oertel, and H.-J. Schulze,
 Phys. Lett. {\bf B562}, 153 (2003).
\bibitem{cdm}
% HYBRID STARS WITH THE COLOR DIELECTRIC AND THE MIT BAG MODELS
 C. Maieron, M. Baldo, G. F. Burgio, and H.-J. Schulze,
 Phys. Rev. {\bf D70}, 043010 (2004).

\bibitem{praka}
% COMPOSITION AND STRUCTURE OF PROTONEUTRON STARS
 M. Prakash, I. Bombaci, M. Prakash, P. J. Ellis, J. M. Lattimer,
 and R. Knorren,
 Phys. Rep., {\bf 280}, 1 (1997).

\bibitem{chodos}
% NEW EXTENDED MODEL OF HADRONS
 A. Chodos,  R. L. Jaffe, K. Johnson, C. B. Thorn, and V. F. Weisskopf,
 Phys. Rev. {\bf D9}, 3471 (1974).

\bibitem{alford}
% COMPACT STARS WITH COLOR SUPERCONDUCTING QUARK MATTER
 M. Alford and S. Reddy,
 Phys. Rev. {\bf D67}, 074024 (2003).

\bibitem{glen}
% FIRST-ORDER PHASE TRANSITIONS WITH MORE THAN ONE CONSERVED CHARGE:
% CONSEQUENCES FOR NEUTRON STARS
 N. K. Glendenning,
 Phys. Rev. {\bf D46}, 1274 (1992).

\bibitem{mixed}
% CHARGE SCREENING EFFECT IN THE HADRON-QUARK MIXED PHASE
 T. Endo, T. Maruyama, S. Chiba, and T. Tatsumi,
 Prog. Theor. Phys., {\bf 115}, 337 (2006).

%\bibitem{nv}
%% NEUTRON STAR MATTER AT SUB-NUCLEAR DENSITIES
% J. W. Negele and D. Vautherin,
% Nucl. Phys. {\bf A207}, 298 (1973).
\bibitem{bps}
% THE GROUND STATE OF MATTER AT HIGH DENSITIES:
% EQUATION OF STATE AND STELLAR MODELS
 G. Baym, C. Pethick, and D. Sutherland,
 Astrophys. J. {\bf 170}, 299 (1971).
\bibitem{fmt}
% EQUATIONS OF STATE OF ELEMENTS BASED ON THE GENERALIZED FERMI-THOMAS THEORY
 R. Feynman, F. Metropolis, and E. Teller,
 Phys. Rev. {\bf C75}, 1561 (1949).

\bibitem{nicot1}
 O. E. Nicotra,
% Structure of protoneutron stars within a static approach
 arXiv:nucl-th/0607055.
 
\bibitem{lat}
% RAPIDLY ROTATING PULSARS AND THE EQUATION OF STATE
 J. M. Lattimer and F. D. Swesty,
 Nucl.Phys. {\bf A535 }, 331 (1997).

\bibitem{burr}
% THE BIRTH OF NEUTRON STARS
 A. Burrows and J. M. Lattimer,
 Astrophys. J. {\bf 307}, 178 (1986).

\bibitem{strobel}
% PROPERTIES OF NON-ROTATING AND RAPIDLY ROTATING PROTONEUTRON STARS
 K. Strobel, C. Schaab, and M. K. Weigel,
 Astron. Astrophys. {\bf 350}, 497 (1999).

\bibitem{pons}
% EVOLUTION OF PROTO-NEUTRON STARS
 J. A. Pons, S. Reddy, M. Prakash, J. M. Lattimer, and J. A. Miralles,
 Astrophys. J. {\bf 513}, 780 (1999);
% EVOLUTIONARY SEQUENCES OF ROTATING PROTONEUTRON STARS
 L. Villain, J. A. Pons, P. Cerd\'a-Dur\'an, and E. Gourgoulhon,
 Astron. Astrophys. {\bf 418}, 283 (2004).

\bibitem{cooke}
% QUARK-HADRON PHASE TRANSITION IN PROTONEUTRON STARS
 M. Prakash, J. R. Cooke, and J. M. Lattimer,
 Phys. Rev. {\bf D52}, 661 (1995);
% QUARK-HADRON PHASE TRANSITIONS IN YOUNG AND OLD NEUTRON STARS
 A. W. Steiner, M. Prakash, and J. M. Lattimer,
 Phys. Lett. {\bf B486}, 239 (2000);
% EVOLUTION OF PROTO-NEUTRON STARS WITH QUARKS
 J. A. Pons, A. W. Steiner, M. Prakash, and J. M. Lattimer,
 Phys. Rev. Lett. {\bf 86}, 5223 (2001).
\bibitem{lugon}
% EFFECT OF TRAPPED NEUTRINOS IN THE HADRON MATTER TO QUARK MATTER TRANSITION
 G. Lugones and O. G. Benvenuto,
 Phys. Rev. {\bf D58}, 083001 (1998).
\bibitem{nemes}
% NEUTRINO TRAPPING AND HYBRID PROTONEUTRON STAR FORMATION
 S. Epsztein Grynberg, M. C. Nemes, H. Rodrigues, M. Chiapparini,
 S. B. Duarte, A. H. Blin, and B. Hiller,
 Phys. Rev. {\bf D62}, 123003 (2000).

\bibitem{kaaret}
% STRONG-FIELD GENERAL RELATIVITY AND QUASI-PERIODIC OSCILLATIONS
% IN X-RAY BINARIES
 P. Kaaret, E. Ford, and K. Chen,
 Astrophys. J. Lett. {\bf 480}, L27 (1997);
% CORRELATION BETWEEN ENERGY SPECTRAL STATES AND FAST TIME VARIABILITY
% AND FURTHER EVIDENCE FOR THE MARGINALLY STABLE ORBIT IN 4U 1820-30
 W. Zhang, A. P. Smale, T. E. Strohmayer, and J. H. Swank,
 Astrophys. J. Lett. {\bf 500}, L171 (1998);
% A 2.1 MSOLAR PULSAR MEASURED BY RELATIVISTIC ORBITAL DECAY
 D. J. Nice, E. M. Splaver, I. H. Stairs, O. L\"ohmer, A. Jessner, M. Kramer,
 and J. M. Cordes,
 Astrophys. J. {\bf 634}, 1242 (2005).

\end{thebibliography}
\end{document}